\documentclass[journal]{IEEEtran}

\usepackage{cite}
\usepackage[pdftex]{graphicx}
\graphicspath{{./figures/}}
\DeclareGraphicsExtensions{.pdf, .jpg, .jpeg}
\usepackage{amsmath, amssymb, amsfonts}
\usepackage{algorithm, algorithmic}
\usepackage{xcolor}
\usepackage{siunitx} 
\usepackage{bm}

\newcommand{\specialcell}[2][c]{%
  \begin{tabular}[#1]{@{}c@{}}#2\end{tabular}}

\makeatletter
\let\NAT@parse\undefined
\makeatother

\usepackage{hyperref}
\hypersetup{hidelinks}
\usepackage{textcomp}

\title{Collaborative Drill Alignment in Surgical Robotics}

\author{Daniel Larby, \IEEEmembership{Member, IEEE}, Joshua Kershaw, Matthew Allen, and Fulvio Forni, \IEEEmembership{Senior Member, IEEE} 
\thanks{Manuscript received Jan 28, 2024; revised XXXXXX XX, XXXX.} %
\thanks{This work was partially supported by the project ‘Robot Assisted Veterinary Orthopaedics’ funded by
the School of Technology Seed Fund of the University of Cambridge, UK. For the purpose of open access, the author has applied a Creative Commons Attribution (CC BY) licence to any Author Accepted Manuscript version arising.} %
\thanks{D. Larby and F. Forni were both with the Control Lab, Department of Engineering, University of Cambridge, CB2 1PZ, UK. However, Daniel has since moved to Swan Endosurgical, 155 Cambridge Science Park, Cambridge CB4 0GN (email: dan\_larby@hotmail.co.uk; f.forni@eng.cam.ac.uk)} %
\thanks{J. Kershaw and M. Allen are with the Department of Veterinary Medicine, University of Cambridge, CB3 0ES, UK. (email: jtk32@cam.ac.uk; mja1000@cam.ac.uk)}}

\begin{document}

\def\libname{VMRobotControl.jl}
\def\julia{JULIA}
\def\RSON{.rson}
\def\vm{virtual mechanism}
\def\vmc{virtual mechanism control}
\newtheorem{remark}{Remark}

\def\pprobe{\bm{p}^p_{\mathrm{probe}}}

\def\ptipdrill{\bm{p}^d_{\mathrm{tip}}}
\def\adrill{\bm{a}_{\mathrm{bit}}^d}
\def\pbasedrill{\bm{p}^d{\mathrm{base}}}

\def\ptipee{\bm{p}^e_{\mathrm{tip}}}
\def\aee{\bm{a}_{\mathrm{bit}}^e}
\def\pbaseee{{\bm{p}^e{\mathrm{base}}}}

\maketitle

\begin{abstract}
Robotic assistance allows surgeries to be reliably and accurately executed while still under direct supervision of the surgeon, combining the strengths of robotic technology with the surgeon's expertise.
This paper describes a robotic system designed to assist in surgical procedures by implementing a virtual drill guide. 
The system integrates virtual-fixture functionality using a novel virtual-mechanism controller with additional visual feedback.
The controller constrains the tool to the desired axis, while allowing axial motion to remain under the surgeon's control.
Compared to prior virtual-fixture approaches---which primarily perform pure energy-shaping and damping injection with linear springs and dampers--our controller uses a virtual prismatic joint to which the robot is constrained by nonlinear springs, allowing us to easily shape the dynamics of the system.
We detail the calibration procedures required to achieve sufficient precision, and describe the implementation of the controller.
We apply this system to a veterinary procedure: drilling for transcondylar screw placement in dogs.
The results of the trials on 3D-printed bone models demonstrate sufficient precision to perform the procedure and suggest improved angular accuracy and reduced exit translation errors compared to patient specific guides (PSG). 
Discussion and future improvements follow.
\end{abstract}

\section{Introduction}

\IEEEPARstart{R}{obotic} surgery has many potential advantages for treatment of humeral intracondylar fissure in dogs: these include enhanced precision, accuracy and reliability.
We propose a robotic system to assist with drilling in preparation for transcondylar screw placement \cite{Moores2021}.
The novelty of our approach is to cast the problem into the setting of collaborative robotics: replacing the physical guide with a virtual one, combining the skills of the surgeon with the precision of the robot, and implementing this in an interactive way, not obscured by teleoperation or taken out of the surgeons hands by automation.
We ask: can a mechanical drill guide be replaced by a virtual-drill guide, enforced by the robot?
How can a controller be designed to implement such a behaviour?
Can we show that the performance/accuracy of this system is sufficient and compares favourably to other methods?

We will contrast our approach with a state-of-the-art assistive technology: 3D printed Patient-Specific-Guides (PSGs).
A PSG is made by taking a template 3D model of a guide and subtracting the shape of the bone from it (taken from the CT scan), so that the surface of the PSG conforms to the bone. A clearance hole is then added to constrain the motion of the drill bit along an axial direction \cite{Easter2020}.
Compared to a PSG, a virtual guide offers several benefits.
A PSG must be prepared for each individual surgery, increasing the lead time.
PSGs require more soft tissue to be cleared so that the drill guide can sit cleanly on the bone.
In contrast, a robotic approach does not require a custom part to be made and drilling can be performed minimally invasively, possibly decreasing the risk of infection.

\begin{figure}[htbp]
    \centering \includegraphics[page=2,width=\columnwidth]{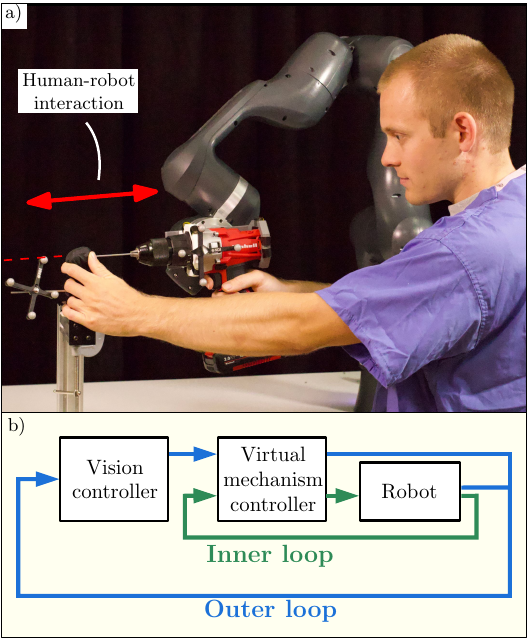}
    \caption{ a) Experimental setup in position for drilling, from the perspective of the vision sensor.
    b) Block diagram of the control architecture, showing inner and outer control loops. More details in Section \ref{sec:VirtualDrillGuide}.}
    \label{fig:ControllerImplementation}
\end{figure}

As shown in Figure \ref{fig:ControllerImplementation}, our system utilizes a two-feedback approach.
Firstly, a virtual mechanism \cite{Larby2024, Zhang2024, Larby2023, Larby2022b, Pratt1995, Joly1995} forms the fast {($\SI{1000}{Hz}$)} inner-loop robot controller and allows compliant co-manipulation with the surgeon, so that the robot can act as an assistive controller.
Secondly, a low-rate {($\SI{20}{Hz}$)} outer-loop vision feedback controller adapts the virtual mechanism based on it's more accurate but slower measurements, to improve accuracy to the required level.
The robot performs the same function as a physical drill guide, `automating' the task of maintaining an accurate position/orientation, while leaving axial motion in the hands of the surgeon.

The virtual mechanism that defines the control action in the inner loop is designed using simple physical analogies that make the purpose of the controller easy to understand. The virtual mechanism takes the form of a virtual drill interconnected to the robot arm via nonlinear saturating springs. 
{The maximum force these springs produce is bounded}, improving safety during interaction.
The vision controller in the outer loop updates the position of the virtual drill to match the measured position of the bone. It also  continuously adapts the offset of the saturating springs in response to visually measured error.
The combined inner- and outer-loop actions lead to a torque reference for the torque controlled joints of the robot. 
The motion of the robot is constrained by the virtual mechanism while preserving backdrivability in the null-space of the constraints, so the redundancy of the robot can be exploited to change the pose during operation.

\IEEEpubidadjcol

Compared with existing work in surgical robotics and virtual fixtures, we emphasize several novelties.
There exist commercial systems for spinal screw placement \cite{DSouza2019, Lieberman2020, Lieberman2006, Smith2021}, but these simply position a drill guide, which means there is no ability to shape dynamics along the drilling axis.
Another difference is that in \cite{Lieberman2006, Lieberman2020} the custom miniature-robot is rigidly fixed to the spine to avoid relative motion---we instead track the bone with the vision system and use a standard serial robot arm.
Other orthopaedic surgical systems which use a virtual fixture approach focus on knee and hip replacement/arthroplasty rather than screw placement \cite{Davies2006a, Lopez2013, Davies2007}.
Compared to virtual-fixture/active constraints works \cite{Pezzementi2007, Bowyer2014, Selvaggio2018} we emphasize the novelty of the nonlinear springs in our approach, which are used to perform energy shaping to saturate the control effort; the novelty of the simulated/proxy virtual drill, which allows shaping of the dynamics along the direction of motion; and the novelty of the integration the vision system into the virtual mechanism.

In Section \ref{sec:Overview} we provide an overview of the functionality of the whole system and it's constituent parts.
Section \ref{sec:CalibrationMethods} discusses in the mathematical methods used for system calibration while Section \ref{sec:CalibrationProcedure} discusses the calibration procedure.
Section \ref{sec:VirtualDrillGuide} explains the controller.
We feature results and discussion from a trial of 16 3D printed bones in Section \ref{sec:Results} and future directions in Section \ref{sec:FutureDirections}.

\subsection{Notation}
Bold typeface denote vectors (e.g. $\bm{p}$). The letter $T$ is used for rigid transforms. Uppercase letters are used for matrices unless otherwise indicated. We will use the notation $T^{ab}$ to represent the rigid transform from frame $b$ to frame $a$, and superscript to denote the frame of reference an object is represented in e.g. $\bm{p}^b$ is the representation of $\bm{p}$ in frame $b$. The frame abbreviations used throughout the paper are:
{
\begin{table}[ht]
    \small
    \begin{center}
    \begin{tabular}{| r | c |}
        \hline
         Frame & Abrv. \\
         \hline
         Vision Sensor & v \\
         Robot & r \\
         CT Scan & s \\
         Probe & p \\
         \hline
    \end{tabular}
    \hspace{1mm}
    \begin{tabular}{| r | c |}
        \hline
         Frame & Abrv. \\
         \hline
         Bone tracker & b \\
         Measuring rod & m \\
         Drill tracker & d \\
         Robot End-Effector & e \\
         \hline
    \end{tabular}
    \vspace{1mm}
    \caption{Frame abbreviations.}
    \label{tab:frames}
    \end{center}
\end{table}
}

Equation \eqref{eq:T} shows how to apply transform $T^{ab}$ to point $\bm{p}^b$, transforming it to frame $a$. 
\begin{align}
    \bm{p}^a &= T^{ab}(\bm{p}^b) \label{eq:T}
\end{align}
Sometimes we represent transforms  as a translation vector $\bm{o}^{ab}$ and a rotation matrix $R^{ab}$, as shown in.\eqref{eq:o_and_R}.
\begin{align}
    \bm{p}^a &= R^{ab}\bm{p}^b + \bm{o}^{ab} \label{eq:o_and_R}
\end{align}

\section{Overview}
\label{sec:Overview}

The goal of our robot-surgeon collaborative system is to drill a hole in the patient's bone, at the target entry/exit location.
To do this, we must know the planned entry and exit points and the location of the bone relative to the robot.
As the bone may move during the operation, we must track the bone, which is achieved by drilling an Ellis pin into the bone with a rigid body tracker attached (as seen in  Figure \ref{fig:ControllerImplementation} and later in Figure \ref{fig:BoneCovered}), which can be used to sense the position/orientation and thus determine a rigid-body transform between the tracker and the vision sensor.
{This is a common and practical procedure for navigated surgery \cite{Kamara2017}.}

\begin{figure}[!t]
    \centering
    \includegraphics{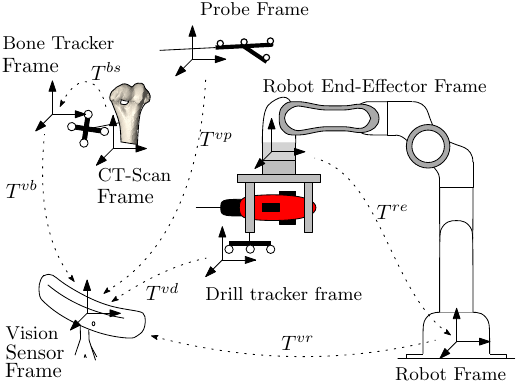}
    \caption{Reference frames and the transformations between them.}
    \label{fig:Transforms}
\end{figure}

Thus, we are challenged to work with several reference frames, as highlighted by Figure \ref{fig:Transforms}. 
We must be able to relate the CT scan to the bone-tracker, the bone tracker to the camera, and the camera to the drill (mounted on the robot).
This is done via several calibration steps, to ensure the highest possible accuracy.

To determine the position of the bone-tracker relative to the CT scan of the bone, $T^{bs}$, the surgeon measures several landmark points on the bone, as designated in the CT-scan by the surgeon, with a probe. The details of this procedure are in Section \ref{sec:bone_registration}.
Then we can use the measurement of bone-tracker $T^{vb}$ to relate the CT scan to the vision sensor frame.

We must then find the correspondence between the vision sensor frame and the drill, which can be done in two ways. The first way is to perform a hand-eye registration to determine the transform between the robot frame and the vision sensor, $T^{vr}$. 
Then the robot kinematics can be used to relate the robot frame to the drill end-effector frame (in which the drill is fixed).
The second way is to mount a tracker to the drill, and measuring it with the vision sensor directly. 
We do both for a couple of reasons: the accuracy of the vision sensor is better than the robot\footnote{While the repeatability of the robot---its end-effector position accuracy when returning to the same joint angles---is sub-millimetre, we found that accuracy of its end-effector positioning is in fact on the order of millimetres}, so by visually measuring both the bone and the drill we ensure the best possible performance. 
However, the vision sensors are susceptible to occlusion and provides measurement at a lower rate than the robot.
The details of how we use these mixed feedback signals are explained in Section \ref{sec:VirtualDrillGuide}.

We must also, of course, calibrate the probe and the drill.
We perform a calibration to determine the position of the tip of the probe, $\pprobe$, to attain the highest accuracy for bone registration.
Finally, we perform 4 distinct calibration steps to determine the tip-position and axis-direction of the drill bit relative to the drill-tracker ($\ptipdrill, \adrill$), and the tip-position/axis-direction of the drill bit in the robot end-effector frame ($\ptipee, \aee$), so that the vision controller and virtual mechanism controller can each determine the location/orientation of the drill bit.

Once all of the position, axes and transform calibrations are completed, we can initialize the virtual-mechanism controller in the correct position for drilling, and use the visual measurements of the bone/drill positions to adapt the controller via the outer feedback loop.

\section{Calibration methods}
\label{sec:CalibrationMethods}

The success of the system depends upon the ability to accurately position and orient the drill bit. 
There are many possible ways to calibrate the transforms/parameters of interest. 
We choose to we use methods that rely upon taking many redundant measurements and performing a best-fit, to reduce the effects of noise.
Due to the importance of calibration, we cover the mathematical details of these best-fit methods here.

\subsection{Fixed point calibration} \label{sec:FixedPointCalibration}

\begin{figure}[!t]
    \centering
    \includegraphics{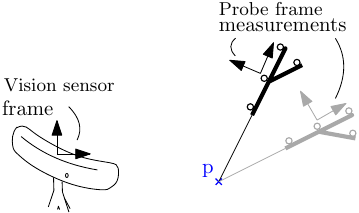}
    \caption{Probe calibration measurements: the probe reference frame is shown in two configurations, pivoting around fixed point $p$.}
    \label{fig:ProbeCalibration}
\end{figure}

We need to be able to find the position of the probe tip and the tip of the drill bit within their reference frame, so the first type of calibration we will perform allows us accurately determine the position of such a point.
Consider the task of identifying the probe tip $\pprobe$, based upon measurements transform, $T^{vp}$.
We will take many measurements of $T^{vp}$ while the probe is pivoted around its tip, assuming the sensor is stationary, at a point in the sensor frame which we will call $\bm{p}^v$, as illustrated in Figure \ref{fig:ProbeCalibration}.
With these measurements we can formulate a least squares minimisation to find $\pprobe$ and $\bm{p}^v$ simultaneously.

In general, to identify point $\bm{p}^b$ from a measurements of $T^{ab}$, we take many measurements with the point $\bm{p}^b$ at the same place in frame $a$ (i.e. at the point we call $\bm{p}^a$), but moving the body $b$ to various orientations (e.g., in Figure \ref{fig:ProbeCalibration} the probe would be the body b).
The least squares minimization is built by rearranging \eqref{eq:o_and_R} to form
\begin{align}
    \label{eq:fixed_point_calibration_single}
    - R^{ab} \bm{p}^{b} + \bm{p}^a &= \bm{o}^{ab}.
\end{align}
This form is linear in $\bm{p}^a$ and $\bm{p}^b$ on the left side so by concatenating $N$ instances of equation \eqref{eq:fixed_point_calibration_single} for different measurements of $T^{ab}$: $T^{ab}_i$ for $i=1\dots N$, we compose them into a standard least squares problem $A\bm{x}=\bm{y}$
\begin{equation}
    \label{eq:fixed_point_calibration}
    \underbrace{
        \begin{bmatrix}
            -R^{ab}_1 & I_{3\times 3} \\
            -R^{ab}_2 & I_{3\times 3} \\
            \vdots & \vdots \\
            -R^{ab}_N & I_{3\times 3}
        \end{bmatrix}    
        \begin{bmatrix}
            \bm{p}^b \\
            \bm{p}^a
        \end{bmatrix}
    }_{A\bm{x}}
    =
    \underbrace{
        \begin{bmatrix}
            \bm{o}^{ab}_1 \\
            \bm{o}^{ab}_2 \\
            \vdots \\
            \bm{o}^{ab}_N
        \end{bmatrix}
    }_{\bm{y}}
\end{equation}
which can be solved as $\bm{x}^* =(A^TA)^{-1}A^T\bm{y}$, where the first and last 3 elements of $\bm{x}^*$ form our estimates of $\bm{p}^b$ and $\bm{p}^a$ respectively.
This solution $\bm{x}^*$ minimizes cost $\sum_i \bm{e}_i^T \bm{e}_i$, where $\bm{e}_i$ is the error $\bm{e}_i = \bm{p}^a - (R^{ab}_i \bm{p}^b + \bm{o}^{ab}_i)$.
The RMS error \eqref{eq:rms_error} for this calibration can be computed as:
\begin{align}
    \label{eq:rms_error}
    e_{\text{rms}} &= \sqrt{\frac{1}{N} \sum_{1:N} \bm{e}_i^T \bm{e}_i}
\end{align}
and has units of distance, so is reported to the user to determine if the accuracy is sufficient. 

\subsection{Axis calibration} 
\label{sec:AxisCalibration}
We need to be able to determine the direction of the drill bit, relative to the drill-tracker frame and the end-effector frame. This second calibration procedure allows us to identify the direction of an axis, after the position of the tip has been calibrated.
The method we propose is similar to that used in \cite{Yang2023} to identify a plane, based on the singular value decomposition (SVD) of a set of measured points.

Consider the moving frame $b$, and a known point, fixed in frame $b$: $\bm{p}^b$.
We take $N$ measurements of $T^{ab}$, under the assumption that
the relative motion between frames $a$ and $b$ is constrained to translation along and rotation around the axis $\bm{a}$, rigidly attached to frame $b$ and passing through $\bm{p}^b$.
For each measurement $T^{ab}_i$, a position $\bm{p}^a_i=T^{ab}_i\bm{p}^b$ is stored.
We wish to compute an explicit representation of $\bm{a}^b$ .
To do this, we first compute the centroid of points $\bm{p}^a_i$:
\[
    \bm{\mu}^a = \frac{1}{N}\sum_{i\in{1:N}}(\bm{p}^{a}_i),
\]
which should lie on the axis.
Then, the position of the centroid is transformed back to frame $b$ for each measurement
\[
    \bm{\mu}^b_i = T_i^{ba}(\bm{\mu}_a)
\]
The SVD is taken as shown in \eqref{eq:svd1}, \eqref{eq:svd2}, and the column of $V$ corresponding to the largest singular value gives us an estimate for the axis $\bm{a}^b$
\begin{eqnarray}
    \label{eq:svd1}
    \tilde{\bm{\mu}}^b = \frac{1}{N}\sum_{i\in1:N} \bm{\mu}_i^{b}, 
    &
    M = \begin{bmatrix}
        (\bm{\mu}_1^{b} - \tilde{\bm{\mu}}^b)^T \\ (\bm{\mu}_2^{b} - \tilde{\bm{\mu}}^b)^T \\ \vdots \\ (\bm{\mu}_N^{b} - \tilde{\bm{\mu}}^b)^T
    \end{bmatrix} \\
    \label{eq:svd2}
    U\Sigma V^T = M, 
    &
    \bm{a}^b = V \begin{bmatrix}
        1 \\ 0 \\ 0
    \end{bmatrix}
\end{eqnarray}
Note that the direction of the vector returned by the SVD may be opposite to the intended direction, therefore one may need to flip the result by negation $-\bm{a}^b$.
We report an RMS error {computed \color{blue}} using equation \eqref{eq:rms_error} with error 
\begin{equation}
\label{eq:axis_rms_err}
\bm{e}_i= (\bm{\mu}_i^b - \bm{p}^b) - ((\bm{\mu}_i^b - \bm{p}^b)\cdot \bm{a}^b) \bm{a}^b,
\end{equation}
which is the displacement from the closest point on a line with direction $\bm{a}^b$ passing through point $\bm{p}^b$ to $\bm{\mu}_i^b$.

\subsection{Transform Registration} \label{sec:TransformRegistration}

A final calibration method is used to determine the transformation between two reference frames, given matching measurements of points.
Consider two ordered sets of $N$ corresponding points $\mathcal{A}=\{\bm{p}_1^a, \bm{p}_2^a\dots\}$ and $\mathcal{B}=\{\bm{p}_1^b, \bm{p}_2^b,\dots\}$.
With these, Arun's method \cite{Arun1987} can be used estimate the transform $T^{ab}$.
This method uses the singular value decomposition to find the best-fit rotation, and we report to the user the RMS error \eqref{eq:rms_error} with error
\[
    \bm{e}_i = T^{ab}(\bm{p}_i^b) - \bm{p}_i^a.
\]

\section{Calibration Procedures}
\label{sec:CalibrationProcedure}

\subsection{Measuring rigid transformations}

\begin{figure}
    \centering
    \includegraphics[width=\columnwidth]{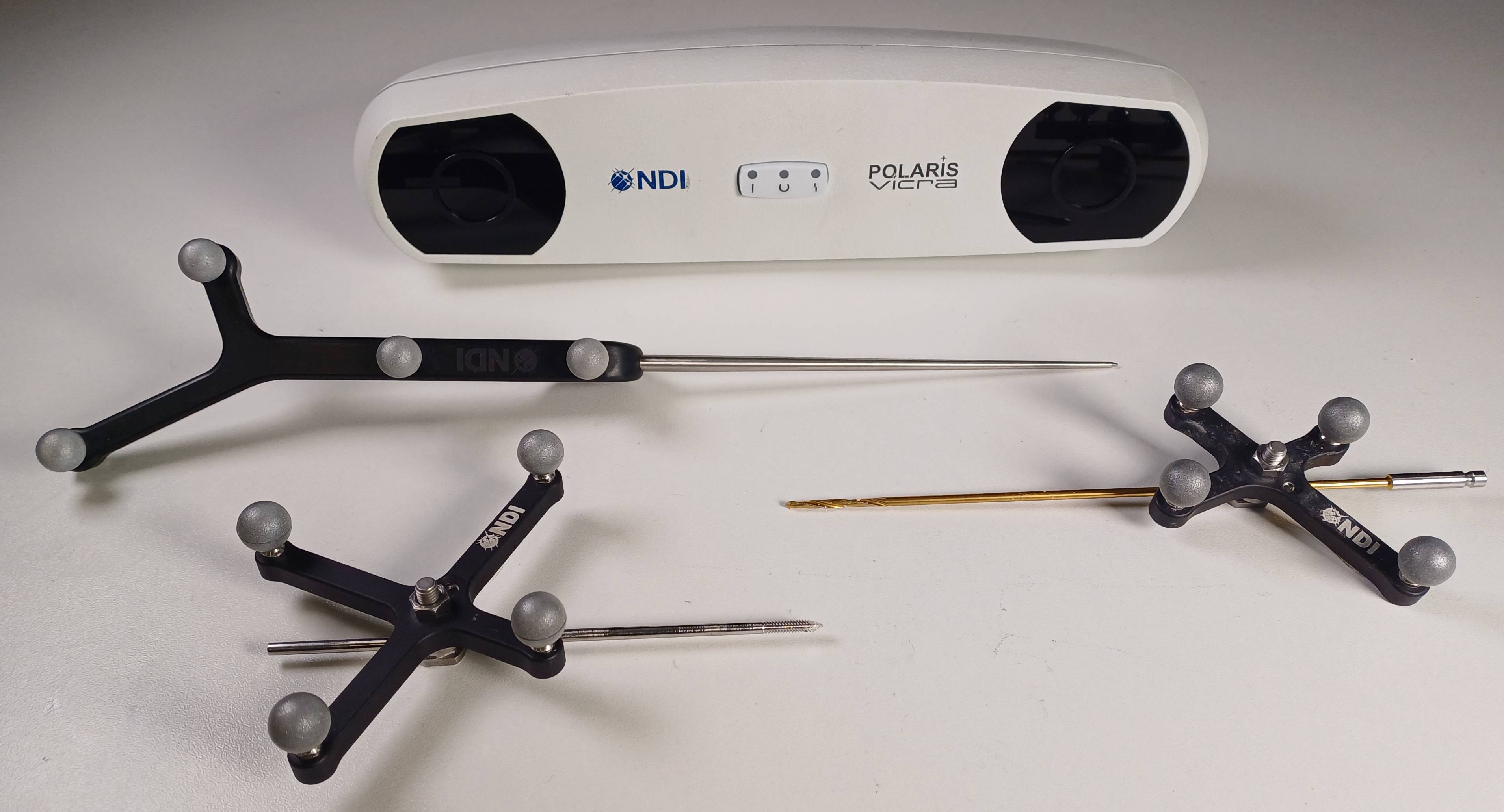}
    \caption{Descending: NDI Polaris Vicra sensor; passive probe; tracked drill bit; and tracked Ellis pin.}
    \label{fig:VADAR_setup}
\end{figure}

For the vision sensor, we use the NDI Polaris Vicra, along with several rigid body tracking accessories shown in Figure \ref{fig:VADAR_setup}, which claims a tracking accuracy 95\% and a confidence interval of $\SI{0.5}{mm}$.
The Vicra system (Figure \ref{fig:VADAR_setup}) communicates to the computer via a USB-serial connection, using the SciKit-Surgery NDI tracker Python library \cite{SciKitSurgery} and then forwarding this information to the main controller via a UDP interface.

The Vicra system can return a rigid body transform for each object currently in view.
The frame designations for all the reference frames used in this paper are shown in Table \ref{tab:frames}.

\subsection{Probe calibration}
\label{sec:ProbeCalibration}

Probe calibration is achieved using the method in Section \ref{sec:FixedPointCalibration}.
The probe is slowly pivoted around a point, as illustrated in Figure \ref{fig:ProbeCalibration}, while we continuously measure $T^{vp}$ at the framerate of the Vicra sensor {(20Hz)} over 30 seconds of recording.
The best accuracy is achieved with pristine reflective-markers (no scratches or damage), the Vicra sensor at it's minimum working distance from the probe, and using an indent on a \emph{hard}  surface to pivot around, with which we obtain calibrations achieving $e_{\text{rms}} \in (\SI{0.2}{mm}, \SI{0.4}{mm})$, {calculated using \eqref{eq:rms_error}.}
To ensure that the calibration is valid, measurements should be taken over a significant range of motion of the probe.

Once the calibration is done, the position of the probe-tip in the probe frame, $\pprobe$, is saved so that the probe can be used to accurately measure points for the bone-registration.

\subsection{Drill tip calibration}

The drill mount seen in Figure \ref{fig:ControllerImplementation} rigidly connects the drill to the end-effector of the robot, and it's attached rigid body tracker.
The drill is securely clamped in place with 8 M6 bolts, which allows an off-the-shelf tool to be attached without requiring an accurate CAD model of the drill.
As a result, the exact position of the drill bit is not known, so a calibration step is needed to determine the position and orientation of the drill bit, relative the end-effector frame of the robot and the frame of the rigid-body-tracker.

\begin{figure}[hb]
    \centering
    \includegraphics[]{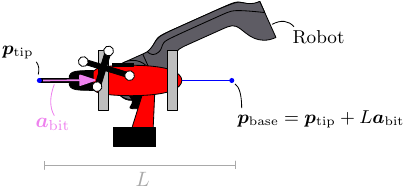}
    \caption{Drill geometry. Calibrated point and axis, $\bm{p}_{\text{tip}}$ and $\bm{a}_{\text{bit}}$ are shown, as well as derived point $\bm{p}_{\text{base}}$.}
    \label{fig:DrillGeometry}
\end{figure}

Drill tip calibration is also achieved using the method in Section \ref{sec:FixedPointCalibration}, but must be performed twice, to determine the position of the tip in the drill tracker frame $\ptipdrill$ (used by the outer-loop vision controller) and in the end-effector frame $\ptipee$ (used by the inner-loop virtual mechanism controller). 
{The point $\bm{p}_\text{tip}$ is illustrated in Figure \ref{fig:DrillGeometry}.}

To determine $\ptipdrill$, the drill assembly is removed from the robot (the clamp and tracker remain attached to the drill), and slowly pivoted around a fixed point while recording $T^{vd}$ for 30 seconds.
We noted that when calibrating with a thin drill bit, bending of the drill bit leads to inaccurate calibrations.
Therefore, we replace the 2.5mm surgical drill bit with an $\SI{8}{mm}$ diameter bit of equal length during this calibration step.

To determine $\ptipee$, the drill-assembly must now be mounted on the robot, but the same $\SI{8}{mm}$ drill bit is used.
The transform from the robot base to robot end-effector frame $T^{er}$ can be determined using the kinematic model of the robot \cite{FrankaDescription} and the reported joint angles.
Pivoting around the drill tip without slipping is difficult with the robot attached, so measurements of $T^{er}$ are taken individually by placing the robot a new position and taking a single measurement, rather than via a continuous recording.

Similar practical concerns are faced as during probe calibration: the vision sensor should be close, an indented hard surface should be used, and a wide range of motion should be explored to achieve a good calibration.

\subsection{Drill axis calibration}

Knowing accurately the direction of the drill is vital to ensuring accurate operation.
The procedure of Section \ref{sec:AxisCalibration} is used to determine the axis in the drill tracker frame $\adrill$ and in the end-effector frame $\aee$.
Firstly, the drill bit is removed and replaced with a $\approx\SI{100}{mm}$ long $\SI{8}{mm}$ diameter steel rod.
A block of wood with an $\SI{8}{mm}$ diameter clearance hole is clamped over the edge of the table.

When calibrating $\adrill$ the drill assembly is removed from the robot and measurements are taken using the drill tip $\ptipdrill$ as the known point on the axis.
Measurements of the transform $T^{vd}$ are taken continuously over 30 seconds, while sliding and rotating the
drill with the rod passing through the clearance hole.

To calibrate of $\aee$ the drill assembly is once again mounted on the robot, using $\ptipee$ as the known point on the axis. 
Again, measurements are taken individually by positioning the robot, then reading it's joint angles and  using the kinematic model to determine $T^{er}$.
{RMS error, calculated as specified at the end of Section \ref{sec:AxisCalibration}, is shown to the user.}

\subsection{Bone registration} 
\label{sec:bone_registration}

\begin{figure}[!t]
    \centering
    \includegraphics[width=\columnwidth]{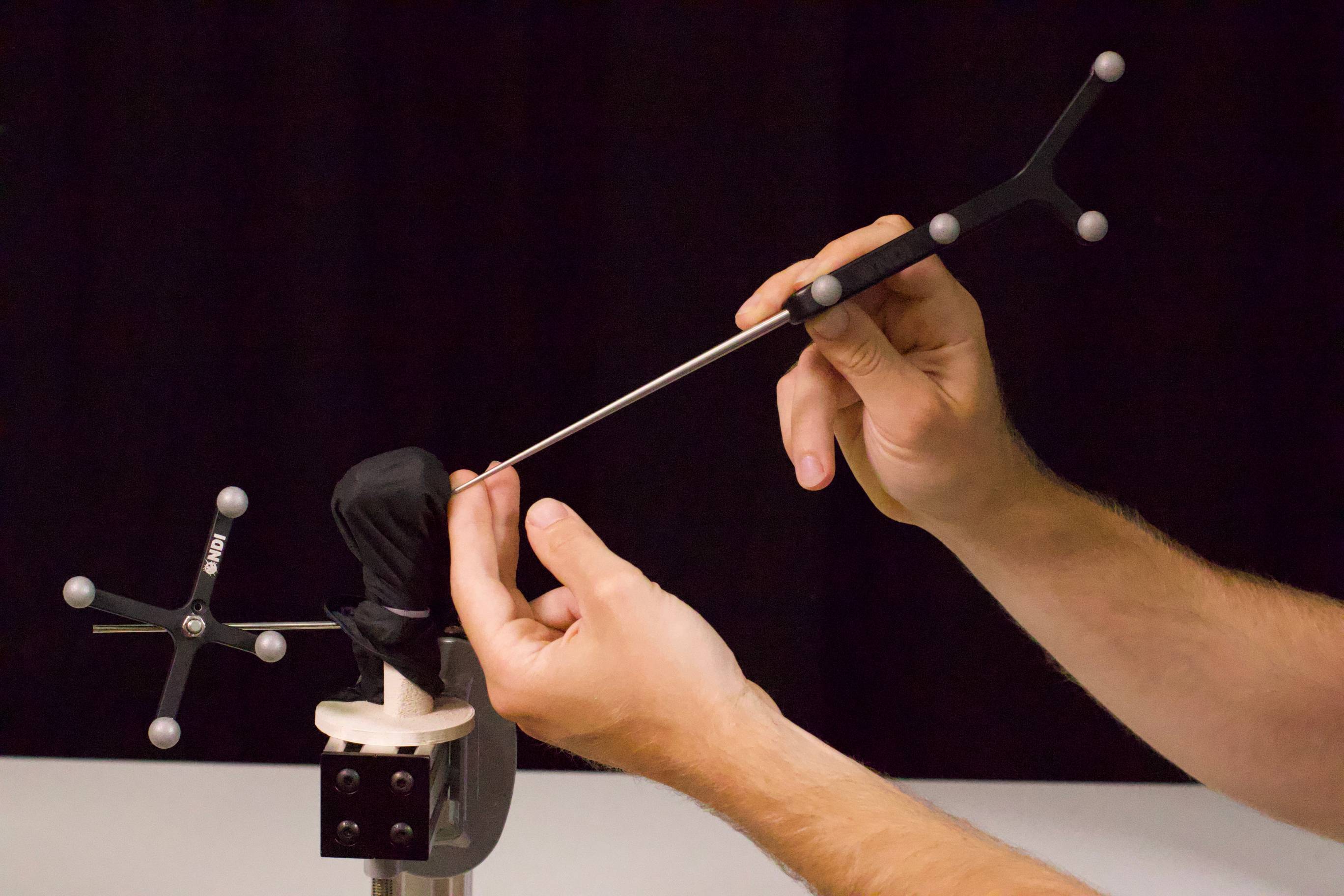}
    \caption{The bone model with attached Ellis pin and bone tracker. A latex covering provides a visual/mechanical barrier between the surgeon and the bone to imitate the hindering effect of skin/fascia. The probe is used to indicate a point on the bone, located by touch.}
    \label{fig:BoneCovered}
\end{figure}

In this section, we explain our procedure to estimate the transform from the CT-scan to the bone-tracker $T^{bs}$.
This allows us to relate the surgical plan to other frames, such as the vision sensor frame or Robot frame.

To do this we use the registration method in Section \ref{sec:TransformRegistration}, for which we must build the two lists of corresponding points $\mathcal{A}$ and $\mathcal{B}$.
Before the procedure, the surgeon annotates on the CT-scan of the bone the desired entry and exit points, and several registration points: features on the bone that can be located by the surgeon by touch, and measured by pointing at them with the probe.
These points are shown in Figure \ref{fig:RegistrationPoints}.

Consider a set of registration points $\{\bm{r}^s_1, \bm{r}^s_2, \dots, \bm{r}^s_N\}$ in the reference frame of the scan.
We take a corresponding set of measurements in the reference frame of the bone, using the probe.
Measurement $i$ of point $\bm{r}_1$ in the bone frame is denoted $\bm{r}^b_{1,i}$ and may be taken by placing the tip of the probe on $\bm{r}_1$ and recording $\bm{r}_{1}^{b} = T^{bv}(T^{vp} (\pprobe))$ using the calibrated probe tip position $\pprobe$, and measured transforms $T^{bv}$ and $T^{vp}$.
If the (simultaneous) measurement of either transform fails, the user is prompted to try again.
To prevent occlusion or change the field of view, it is possible to move the vision sensor during the registration process, because all measurements are stored relative to the bone-tracker frame.

Let $\mathcal{A}$ be the ordered set containing \emph{all} of the measurements.
Then $\mathcal{B}$ is constructed to contain the corresponding reference points.
For example, if we have two measurements of $\bm{r}_1$ and one of $\bm{r}_2$ we obtain $\mathcal{A}=\{\bm{r}^b_{1,1}, \bm{r}^b_{1,2}, \bm{r}^b_{2,1}\}$, $\mathcal{B}=\{\bm{r}^s_1, \bm{r}^s_1, \bm{r}^s_2\}$.
Then, $T^{bs}$ is computed as described in \ref{sec:TransformRegistration}.
A similar point-matching approach is used in other navigated surgeries \cite{Nakamura2009, Audenaert2012} although details of the implementation are not provided.

When performing the registration, the bone-model is covered by a sheet of latex material to prevent the surgeon from using visual cues to locate the registration points, as shown in Figure \ref{fig:BoneCovered}.
If at least 3 points have been measured, the current RMS error of the registration, computed as specified at the end of Section \ref{sec:TransformRegistration}, is printed.
The operator is presented with a live view of the tip of the probe relative to the CT-scan, using the current best-fit transform.
The operator can cycle through the points $\bm{r}_1$ to $\bm{r}_7$, take measurements, and delete the most recent measurement.
If either the bone or probe cannot currently be seen by the vision sensor, the measurement will fail, and the surgeon will be prompted to repeat the measurement.

The final registration procedure is summarized below: 
\begin{enumerate}
    \item Move the probe to point $\bm{r}_1$
    \item Take 5 measurements, with a slightly different probe orientation each time. A measurement may be removed and repeated if the reported error has a large peak (e.g. due to a spurious measurement, the probe slipping...) 
    \item Move to the next point and repeats the above steps for $\bm{r}_2$ to $\bm{r}_7$
    \item Repeat the above steps 2 more times, such that 15 measurement in total are taken for each registration point, for 105 total measurements.
    \item Validate the registration by moving the probe over the bone surface and viewing the live view.
\end{enumerate}
Results of the bone-registration are discussed in Section \ref{sec:Results}.

\subsection{Hand-eye registration}
\label{sec:RobotCameraCalibration}

The final step of setup is the hand-eye registration: to determine the transform from the robot to the vision sensor frame $T^{vr}$.
This is the last to be performed, after all other calibration steps and bone registration.
From this point onwards, the vision sensor must not be re-positioned.

We use the method of Section \ref{sec:TransformRegistration} with two sets of corresponding points in the vision sensor frame and robot frame measured using the drill tip:
\begin{align}
    \bm{p}^v_{\text{tip}} &= T^{vd}\ptipdrill \\
    \bm{p}^r_{\text{tip}} &= T^{re}\ptipee
\end{align}
Transforms $T^{re}$ and $T^{vd}$ are taken simultaneously from the robot and vision sensor, and the calibrated tip positions $\ptipdrill$ and $\ptipee$ are used.

We typically take 10 measurements in a region around the bone, to form the sets $\mathcal{A}$ and $\mathcal{B}$. Our results show that we achieve an RMS error $e_\text{rm} \in (\SI{3}{mm}, \SI{6}{mm})$.
We believe most of this error is due to inaccuracies in the kinematic model of the robot, due to manufacturing tolerances. This error is later mitigated by the outer vision feedback loop.

\section{Control Design}
\label{sec:VirtualDrillGuide}

\subsection{Structure of inner-loop drill controller}

\begin{figure}
    \centering
    \includegraphics[page=1]{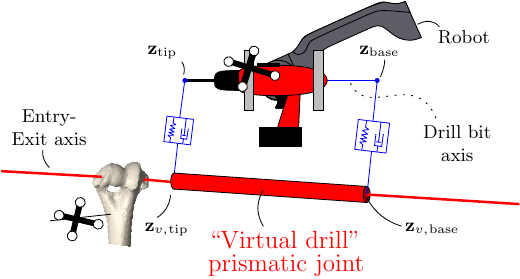}
    \caption{Illustration of the ``Virtual drill'' implementation. A simulated prismatic/sliding joint (red) lies on the entry-exit axis, and is tied to the robot by a pair of spring-dampers (blue).}
    \label{fig:VirtualDrill}
\end{figure}

We design the inner-loop controller using the principles of `design by interconnection' and passivity-based control \cite{Duindam2009,Ortega1998,Ortega2001,Ortega2004}.
In particular we use virtual mechanisms as a design tool to create the robot controller, by reasoning about interconnecting the robot with a mechanical system of components such as joints, springs, and dampers \cite{Larby2024}.
This approach, close to the philosophy of virtual model control \cite{Pratt1995b},  can be seen as a way to perform energy shaping and damping injection, or as a way to design and implement a guidance virtual fixture.  
Rather than considering potential energy functions and guidance-force-fields, we reason about virtual springs interconnected with the robot and the virtual mechanism.

The virtual mechanism approach, like energy-shaping and damping injection, results in a passive controller \cite{Larby2024, Larby2023}.
Passive controllers are particularly well suited to human-robot interaction, as they are stable even when exposed to external forces from another passive system \cite{vanderSchaft2020}.
While the resulting implementation could be considered as an energy-shaping approach, the mechanical interpretation aids with construction and modification of the design.
The surgeon, who will manually drive the system to perform the operation, now has a powerful visual metaphor for the controller which could speed up familiarization with the system (possibly, it also allows the surgeon to contribute to the iterative improvement of the system).
For further detail on virtual mechanism control, see our previous work \cite{Larby2024,Larby2022b,Zhang2024}, our Julia library for constructing, simulating and controlling with virtual mechanisms \cite{VMRobotControl}, the early work of \cite{Pratt1995b, Joly1995}, or more recent work related to virtual fixtures \cite{Zafer2007, Raiola2018}.

The controller is illustrated in Figure \ref{fig:VirtualDrill}.
The virtual mechanism features a virtual drill with a single degree of freedom: a prismatic joint that allows motion along the axis from the desired entry point to the desired exit point, expressed in the robot frame.
It is given an inertance (mass unaffected by gravity) of $m_v$. 
The virtual drill is simulated in real time as part of the controller, using the Euler-Lagrange equations. 
Inertance, $m_v$, and the viscous friction coefficient of the joint, $b_v$, can be used to shape the dynamics along the direction of drilling, by resisting acceleration/velocity.
{The dynamics are specified in \eqref{eq:virtual_drill}.}

Two virtual spring dampers connect the virtual drill to the robot, each with a saturating spring and linear damper.
One is connected at the calibrated location of the drill-bit-tip, to control the position of the drill, with stiffness $k_{\text{tip}}$, spring-saturation force $\sigma_{\text{tip}}$, and damping coefficient $b_{\text{tip}}$ .
The second spring-damper's purpose it to ensure the drill's axis is aligned with the virtual-drill, so is primarily concerned with the \emph{orientation} of the drill.
As such, we choose to connect this spring damper (with parameters $k_{\text{base}}$, $b_{\text{base}}$, $\sigma_{\text{base}}$) a large distance, $L$, from the drill tip, along the drill-bit axis, so that a smaller force is required to rotate the drill.
The springs ensure that (under no external forces) the equilibrium position of the robot/virtual drill system is such that the drill-bit is aligned with the entry-exit axis.

\begin{figure}[t]
    \centering
    \includegraphics{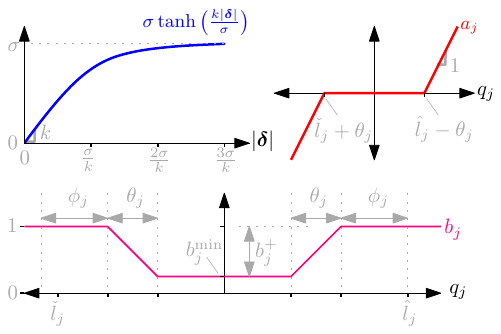}
    \caption{Functions related to virtual spring and damper components.
    }
    \label{fig:SpringDamperCharacteristics}
\end{figure}

The saturating springs have force characteristic 
\begin{align}
    \label{eq:Fspring}
    \bm{f}_{\text{spring}}(\bm{\delta}) &= \sigma \tanh \left( \frac{k|\bm{\delta}|)}{\sigma} \right) \frac{\bm{\delta}}{|\bm{\delta}|}. 
\end{align}
See also Figure \ref{fig:SpringDamperCharacteristics} (top, left).
Here $\bm{\delta}$ is the extension of the spring-damper.
The force saturates at maximum $\sigma$, preventing excessive forces for large displacements, making the mechanisms safer to interact with.
The $\tanh$ function \eqref{eq:Fspring} has a stiffness $k$ when $|\delta|\approx0$.

The linear dampers satisfy
\begin{align}
    \bm{f}_{\text{damper}}(\dot{\bm{\delta}}) &= b\dot{\bm{\delta}}.
\end{align}

Additionally, for each joint $1 \leq j \leq 7$, a joint-limit-enforcing spring-damper is applied at the upper and lower limits of each joint (alongside a constant linear damping term).
A non-linear spring and gain scheduled damper act as virtual-mechanical-buffers and prevent the robot from reaching its hard joint-limits.
Let $q_j$ be be the $j$th joint angle, $\hat{l}_j$ be the upper joint limit, $\check{l}_j$ be the lower joint limit, $\theta_j$ be the distance from the joint limit at which the spring is fully engaged, and $\phi_j$ an additional distance parameter used by the damper.
The damper starts to engage at distance $\theta_j + \phi_j$ and is at it's maximum damping when $q_j > \hat{l} - \theta_j$ at the upper limit or $q_j < \check{l} + \theta_j$ at the lower limit.
With these parameters defined, the demanded joint torques $\bm{\nu}$ of the virtual spring-damper buffer  are element-wise given by
\begin{subequations}
	\begin{align}
		\label{eq:BufferForce}
		\nu_{j} &= \underbrace{k_j a_j(q_j)}_{\text{spring torque}} + \underbrace{b_j(q_j) \dot{q}_j}_{\text{damper torque}} \\
        \label{eq:BufferSpring}
		a_j(q_j) &= 
		\left\{
			\begin{array}{lr}
				q_j - \hat{l}_j - \theta_j &  q_j >\hat{l}_j - \theta_j \\
				0 & \check{l}_j + \theta_j \le q_j \le \hat{l}_j - \theta_j \\
				q_j - \check{l}_j + \theta_j & q_j < \check{l}_j + \theta_j
			\end{array}
		\right.\\
        \label{eq:BufferDampingCoeff}
		\begin{split}
			b_j(q_j) &= b^{\min}_j + b^+_j \mathrm{sat}_0^1\left(\frac{q_j - (\hat{l}_j-\phi_j- \theta_j)}{\phi_j}\right) + \\
			& \qquad\qquad\qquad b_j^+ \mathrm{sat}_0^1\left(\frac{-q_j + (\check{l}_j+\phi_j+\theta_j)}{\phi_j}\right)
		\end{split}
	\end{align}
\end{subequations}
The shape of $a_j$ and $b_j$ (equations \eqref{eq:BufferSpring} and \eqref{eq:BufferDampingCoeff}) are shown in Figure \ref{fig:SpringDamperCharacteristics} (respectively, top-right and bottom plots).

These virtual buffers prevent the robot reaching its hard joint limits and slow down motion near the limits, which prevents the robot from entering a fault state, and the controller from disengaging. The behaviour at joint limits is intuitive, first slowing upon the damper then encountering the spring which prevents reaching the hard limits through a soft constraint.

\subsection{Implementation of the inner loop controller}

The control action of the virtual mechanism is implemented by deriving 
the forces that the virtual mechanism exerts on the robot at different points
via (real-time) simulation. 
These forces are then mapped into demanded torques for the actuators using 
the kinematics of the robot.

Let $\bm{q}=[\bm{q}_r; q_v]$ be the generalized coordinates of the whole system, robot and virtual drill, and $\bm{u}$ the actuation $\bm{u}=[\bm{u_r}; u_v]$.
Consider any virtual component (spring or damper) attached to the robot at generic point $\bm{z}$. $\bm{z}$ is related to the joint coordinates by the forward kinematics,
denoted by $\bm{z}= h(\bm{q})$. Taking the Jacobian {$J_z$} satisfying $\dot{\bm{z}} = J_z(\bm{q})\dot{\bm{q}} = \frac{\partial h(q)}{ \partial \bm{q}}\dot{\bm{q}}$, the force $\bm{f}$
 that the component exerts on $\bm{z}$ (represented in the robot frame) 
 is mapped into the associated vector of joint torques by
\begin{align}
\label{eq:implementation}
    \bm{u} = J_z(\bm{q})^T \bm{f}.
\end{align}
The effect of the torques \eqref{eq:implementation} on the robot movement 
is equivalent to the effect of the virtual force $\bm{f}$ acting at $\bm{z}$, \cite{Khatib1987, Spong1989}.
The effect of multiple components is modeled as a sum of several inputs of the form \eqref{eq:implementation}.
Gravity compensation $\bm{g}(\bm{q_r})$ is also applied, using the modeled mass / kinematics of the robot and the estimated mass / of center mass of the drill.

To implement this type of control, we need only position/velocity kinematics, which are taken from the URDF provided by Franka-Emika \cite{FrankaDescription} (and our calibration procedures).
The relevant virtual forces and Jacobians are computed using our VMRobotControl robotics library \cite{VMRobotControl}, which also provides an interface to construct virtual mechanism controllers in a structured way.

The resulting behaviour can be summarised as follows:
Consider the robot with nominal dynamics:
\begin{equation}
    \label{eq:robot_dynamics}
    M(\bm{q}_r)\ddot{\bm{q}}_r + C(\bm{q}_r, \dot{\bm{q}}_r) \dot{\bm{q}_r} + \bm{g}(\bm{q}_r) = \bm{u}_r + {\bm{u}_e}
\end{equation}
where $M$, $C$ and $\bm{g}$ are the mass matrix, Coriolis matrix and gravity torque vector respectively.
On the right hand side, $\bm{u}_r$ are the joint torques and $\bm{u}_e$ is torque vector resulting from external interaction forces.
Looking at Figure \ref{fig:VirtualDrill}, the controller is implemented as:
\begin{subequations}
\label{eq:ControlLaw}
\begin{align}
    \bm{\delta}_{\text{tip}} &= \bm{z}_{\text{tip}} - \bm{z}_{v,\text{tip}}\\
    \bm{\delta}_{\text{base}} &= \bm{z}_{\text{base}} - \bm{z}_{v,\text{base}}\\
    \bm{f}_{\text{tip}} &= \bm{f}_\text{spring}(\bm{\delta}_{\text{tip}} + \bm{o}_{\text{tip}}) + \bm{f}_\text{damper}(\dot{\bm{\delta}}_{\text{tip}})\\
    \bm{f}_{\text{base}} &= \bm{f}_\text{spring}(\bm{\delta}_{\text{base}} + \bm{o}_{\text{base}}) + \bm{f}_\text{damper}(\dot{\bm{\delta}}_{\text{base}}) \\
	\label{eq:ControlLawUv}
    u_v &= J_{z_{v,\text{tip}}}^T \bm{f}_{\text{tip}} + J_{z_{v,\text{base}}}^T \bm{f}_{\text{base}}\\
	\label{eq:virtual_drill}
    \ddot{q}_v &= m_v^{-1}(u_v - b_v\dot{q}_v) \\
	\label{eq:ControlLawUr}
    \bm{u}_r &= \bm{g} - \bm{\nu} - (J_{z_\text{tip}}^T \bm{f}_{\text{tip}} + J_{z_\text{base}}^T \bm{f}_{\text{base}}). 
\end{align}
\end{subequations}
The additional offsets $\bm{o}_{\text{tip}}$ and $\bm{o}_{\text{base}}$ will be regulated by the outer-loop vision controller, to improve precision. This is discussed in Section \ref{sec:IncorporatingVisualFeedback}.
The positions $\bm{z}_{\text{tip}}$, $\bm{z}_{v,\text{tip}}$, $\bm{z}_{\text{base}}$, $\bm{z}_{v,\text{base}}$ are always represented in the robot frame of reference. 
The tip/base positions are computed using the calibrated drill bit tip, $\ptipee$ and axis, $\aee$, and the transform from the end-effector frame to the robot frame,  $T^{re}$, computed using the robot kinematics and joint angles:
\begin{align}
\bm{z}_\text{tip}&= \mathbf{p}_{\mathrm{tip}}^r = T^{re}(\ptipee) \\
\bm{z}_{\text{base}}&= \mathbf{p}_{\mathrm{base}}^r= T^{re}(\ptipee + L\aee).
\end{align}
$L$ is the virtual drill length.

\begin{remark}
\label{rm:virtual_drill}
Although this approach overlaps with classical virtual fixtures, it also
shows significant differences. The most relevant is that
the controller action is function of the \emph{geometry} and \emph{dynamics} of
the virtual drill along the axial direction \eqref{eq:virtual_drill}, the latter regulated by inertance and damping parameters. Conceptualizing the design of the controller as the design of a virtual mechanisms also allows the user to introduce intuitive modifications, to fine tune the behaviour of the robot. For instance, adaptations to dynamics along the direction of motion such as end-stops can easily be achieved by attaching additional components to the prismatic virtual mechanism, in a way that is similar to the joint-limit buffers. 
Our approach also overlaps with the classical control by interconnection. However, without the intuition of a virtual drill, it could be challenging to find the mathematical expression of the controller required to address the task. In this sense, we think {of} virtual mechanisms as an intuitive way to address control by interconnection.
\end{remark}

\subsection{Parametrization of the inner-loop controller}

The choice of the controller parameters is a compromise between desired performance and implementation constraints, such as robot torque control bandwidth and sensing noise.
If the desired stiffness/damping are too high, the system may become unstable 
due the control loop's effective bandwidth, constrained by the capabilities of the actuators and sensors:  if the system cannot produce the demanded torque then the virtual mechanism behaviour is not accurately emulated, leading to a potential loss of passivity, and instability {\cite{Pratt1995b, Lawrence1988}}. 
Getting close to this limit presents as a poorly damped oscillation, which could be interpreted as being caused by a pole moving close to the imaginary axis in the linearized dynamics. 

Table \ref{tab:ControllerParams} shows the chosen controller parameters.
Ultimately these choices of parameters are based on trial and experimentation, and tuned using intuitive understanding of the function of each component of the virtual mechanism.
To ensure accuracy at the drill tip, $k_\text{tip}$ is large, but $k_{\text{base}}$ is much smaller, as the virtual drill length $L$ gives a large lever arm.
The same reasoning informs the damping coefficients.
Regarding the dynamics of the virtual drill, inertance $m_v$ is dwarved by the inertia of the robot arm: and $b_v$ is dwarved by the effect of $b_\text{tip}$, which is also connected to the virtual drill. 
Appendix \ref{ap:poles} shows an approximate linear dynamics of the virtual drill, which results in oscillatory but quickly decaying poles.
In this particular instance, 
the practical result is similar to the case where no virtual mass is used (and instead the springs automatically pull towards the closest point on the drill axis).

Improved methods are needed for characterizing these limits and tuning controllers appropriately. 

\begin{table}[htbp]
    \centering
    \begin{tabular}{|c|r|}
        \hline
        $m_v$ & $\SI{1}{kg}$ \\
        $b_v$ & $\SI{0.5}{Ns/m}$ \\
        $L$ & $\SI{2}{m}$ \\
        $k_{\text{tip}}$ & $\SI{4000}{N/m}$ \\
        $k_{\text{base}}$ & $\SI{100}{N/m}$ \\
        $b_{\text{tip}}$ & $\SI{40}{Ns/m}$ \\
        $b_{\text{base}}$ & $\SI{0.2}{Ns/m}$ \\
        $\sigma_{\text{tip}}$ & $\SI{20}{N}$ \\
        $\sigma_{\text{base}}$ & $\SI{3}{N}$ \\
        \hline 
    \end{tabular}
    \hspace{.2cm}
    \begin{tabular}{|r|c|c|c|c|c|}
        \hline
        $j$ & $k_j$             & $b_j^{\min}$                 & $b^+_j$               & $\theta_j$        & $\phi_j$ \\
		\hline
             & $\unit{Nm/rad}$ & \multicolumn{2}{c|}{$\unit{Nms/rad}$}   & \multicolumn{2}{c|}{$\unit{rad}$} \\
        \hline \vspace{-2.5mm} & & & & & \\
        \hline
        1 & 50 & 0.5 & 4    & 0.3   & 0.4  \\
        2 & 50 & 0.5 & 4    & 0.2   & 0.2  \\
        3 & 50 & 0.3 & 4    & 0.2   & 0.2  \\
        4 & 60 & 0.3 & 5    & 0.3   & 0.3  \\
        5 & 35 & 0.2 & 2    & 0.35  & 0.35 \\
        6 & 30 & 0.2 & 1.5  & 0.35  & 0.35 \\
        7 & 30 & 0.1 & 1    & 0.35  & 0.35 \\
        \hline
    \end{tabular}
    \vspace{1mm}
    \caption{Inner-loop controller parameters}
    \label{tab:ControllerParams}
\end{table}

\subsection{Improving precision with the outer-loop vision controller}
\label{sec:IncorporatingVisualFeedback}

The closed loop of robot and virtual drill mechanism deliver a controlled constrained behaviour which {is} intrinsically robust. 
However, precision is affected by the uncertainties of the robot kinematics and bone movements, ultimately leading to insufficient performance in the surgical setting. 
We improve precision through adaptation, based on feedback from the vision sensor. 
This is the goal of the outer control loop.

The first step is to compensate for movements of the bone during the operation.
We can determine the current position of the entry/exit points in the robot frame using the transforms from the CT scan to the bone-tracker $T^{bs}$, the bone-tracker to the vision sensor $T^{vb}$, and the vision sensor to the robot $T^{rv}$:
\begin{equation}
    T^{rs} = T^{rv}(T^{vb}(T^{bs})),
\end{equation}
thus:
\begin{align}
    \bm{p}_{\text{entry}}^r &= T^{rs}(\bm{p}^s_{\text{entry}}) \\
    \bm{p}_{\text{exit}}^r &= T^{rs}(\bm{p}^s_{\text{exit}})
\end{align}
These are low pass filtered 
to reduce noise. At each time-step, the virtual-drill is aligned to the filtered entry/exit axis.

The second step is to improve positioning accuracy.
The robot's kinematics are not sufficiently accurate and, to compound the issue, the uncertainty on the calibrated values $\ptipee$ and $\aee$ (based upon the robots kinematics) are worse than the uncertainty on $\ptipdrill$ and $\adrill$ (which are based upon measurements from the vision sensor).
We overcome this challenge by measuring the location of the end-effector via the vision system.
Then, we correct the position of the robot using these measurements, and the more accurate vision-based calibrations of the drill tip/axis. 
These corrections are applied at a lower rate than the inner loop controller{: at $\SI{20}{Hz}$}.
The inner loop, {operating at $\SI{1000}{Hz}$} is able to react quickly disturbances due to its high bandwidth, but the outer loop is ultimately responsible for the overall accuracy of the system.

The outer-loop feedback from the vision system applies adjusts the offsets $\bm{o}_{\text{tip}}$ and $\bm{o}_{\text{base}}$ in \eqref{eq:Fspring} via integral control from the tip-error and the base-error. 
Each integrator has state $\bm{o}$ which is an offset applied to the corresponding spring as described in \eqref{eq:Fspring}.
The locations of the drill tip/base {in the robot frame} are determined using the calibrated drill tip {position} in the drill-tracker frame, $\ptipdrill$, and the drill axis direction in the same frame, $\adrill$,  as:
\begin{align}
    \bar{\bm{z}}_{\text{tip}} &= T^{rv} (T^{vd}(\ptipdrill)) \\
    \bar{\bm{z}}_{\text{base}} &= T^{rv} (T^{vd}(\ptipdrill + L\adrill))
\end{align}
Where the $\bar{\phantom{z}}$ indicates that these are \emph{measures by the vision system}, rather than computed from joint-angles and robot kinematics.
Thus the position errors
\begin{align}
    \bm{e}^r_{\text{tip}} &= \bm{z}_{v,\text{tip}} - \bar{\bm{z}}_{\text{tip}} \\
    \bm{e}^r_{\text{base}} &= \bm{z}_{v,\text{base}} - \bar{\bm{z}}_{\text{base}}
\end{align}
are integrated to correct spring offsets:
\begin{equation}
    \label{eq:Odot}
    \dot{\bm{o}} = \left\{
    \begin{array}{cc}
        0 & \text{If transform measurement fails} \\
         k_i \bm{e}^r & \text{Otherwise}
    \end{array}  
    \right. .
\end{equation}
To mitigate the risk of instability, the integrator state is clamped: if $\bm{o}$ exceeds a predefined bound in any of its x/y/z components, their values are clamped to a maximum absolute value of $\SI{25}{mm}$, i.e. a $\SI{50}{mm}$ edge cube for the tip error; or for the base $\SI{150}{mm}$ i.e. a $\SI{300}{mm}$ edge cube. 
For  a virtual drill of length $\SI{2.0}{m}$ this corresponds to a maximal correction of $\approx 4.3^{\circ}$ of orientation  in pitch and/or yaw.

An advantage of an integral controller is that if the system is subject to only constant disturbances and is stable, then the equilibrium will have zero error.
If the position of the bone tracker moves more than $\SI{75}{mm}$ or rotates by more than $20^\circ$, the controller is immediately terminated, and the robot stops at its current pose.
As a final mitigation of risk, the use of saturating springs for the virtual mechanism ensures that the surgeon can physically override the controller.

This visual feedback loop ensures that the accuracy of the procedure is not limited by the absolute position accuracy of the robot. 
By directly utilizing visual measurements of the tool position and bone position, we achieve the best possible results, limited only by the accuracy of the optical tracking system and the calibrations.

\begin{remark}
As the vision controller directly modifies the extension of the spring and axis of the virtual drill in a non-physical way, it breaks the passivity of the virtual mechanism controller.
However, these corrections act at a lower speeds on a slower time scale, so the rate of energy injection is small.
If $k_i$ is sufficiently small, 
the time-scale separation guarantees stability.

Tracking a moving object with the robot requires a source of energy, so the outer loop is required to inject energy, breaking passivity. 
A precise characterization of the stability margins of the full system is beyond the scope of the current work. 
However, energy-tank or power-valve approaches may be used to certify the stability of the complete system.
\end{remark}
\section{Passivity of the Inner Loop Controller}
In this section we characterize the  interconnection between virtual mechanisms and robot from a system-theoretic perspective, clarifying how passivity provides robust stability guarantees on the closed loop behavior.
Given the nominal dynamics \eqref{eq:robot_dynamics},
and assuming perfect gravity compensation $- G(\bm{q}_r)$, 
the energy of the robot is given by \cite{Spong1989}.
\begin{align}
E_r & = \frac{1}{2}\bm{q}_r^T M(\bm{q}_r)\bm{q}_r 
\end{align}
and is passive, that is, the rate of change of its stored energy is less than or equal to the sum of power flowing into the system from its actuator and external forces:
\begin{align}
	\label{eq:robotEnergy}
	\dot{E}_r & = \dot{\bm{q}}_r^T \bm{u}_e + \dot{\bm{q}}_r^T \bm{u}_r.
\end{align}
In fact, for the real robot, this identity should be replaced by an inequality $\leq$, due to mechanical dissipations not accounted for in \eqref{eq:robot_dynamics}.
Because the controller was designed as a virtual mechanical system, we can associated to it a mechanical energy as well, of the form
\begin{align}
	\label{eq:vmEnergy}
	\begin{split}
		E_v &= \frac{1}{2}m_v q_v^2 + \sum_j \frac{1}{2}k_j a_j^2	\\
		& \qquad+ E_s(\bm{\delta}_\text{tip} + \bm{o}_\text{tip}) + E_s(\bm{\delta}_\text{base} + \bm{o}_\text{base}).
	\end{split}
\end{align}
Respectively, these terms describe the kinetic energy of the virtual drill, 
the potential energy in the joint limit buffer springs, and the potential energy in the tip and base springs\footnote{Note that $E_v$ is a function of both $q_v$ and $\bm{q}_r$ as the virtual springs are attached between the robot and virtual mechanism.}.
The spring energy $E_s$ is
\begin{equation}
E_s(\bm{\delta}) = \frac{\sigma}{k}\ln\left(\cosh\left(\frac{k|\bm{\delta}|}{\sigma}\right)\right),
\end{equation}
such that its derivative is the same form as the spring force, \eqref{eq:Fspring}.
Taking the derivative, we find
\begin{align}
	\begin{split}
		\dot{E}_v &= -b_v\dot{q}_v^2 + \dot{q}_v u_v + \sum_j k_j \delta_j \dot{q}_j \\
		& \qquad + (\dot{\bm{z}}_\text{tip} + \dot{\bm{o}}_\text{tip})^T \bm{f}_\text{spring}(\bm{z}_\text{tip} + \bm{o}_\text{tip}) \\
	 	& \qquad + (\dot{\bm{z}}_\text{base} + \dot{\bm{o}}_\text{base})^T \bm{f}_\text{spring}(\bm{z}_\text{base} + \bm{o}_\text{base})
	\end{split}
\end{align}
Using the sum of the energies $E = E_r + E_v$ as a storage function for the closed loop system, and factoring in \eqref{eq:ControlLawUr} and \eqref{eq:ControlLawUv}, we get  
\begin{align}
    \label{eq:Edot}
	\begin{split}
		\dot{E} &= \dot{\bm{q}}_r^T \bm{u}_e - b_v\dot{q}_v^2 - |\dot{\bm{\delta}}_\text{tip}|^2 b - |\dot{\bm{\delta}}_\text{base}|^2 b - \sum_j b_j(q_j)\dot{q}_j^2 \\ & \quad + \dot{\bm{o}}_\text{tip}^T \bm{f}_\text{spring}(\bm{\delta}_\text{tip} + \bm{o}_\text{tip}) 
		+ \dot{\bm{o}}_\text{base}^T \bm{f}_\text{spring}(\bm{\delta}_\text{base} + \bm{o}_\text{base}) 
	\end{split} 
\end{align}
Respectively these terms are an energy flow from external forces $\bm{u}_e$, four dissipation terms which are non-positive, corresponding to the damping components of the controller, and two power flows into the system due to changes in offsets $\bm{o}_\text{tip}$ and $\bm{o}_\text{base}$.

Without adjusting the offsets (i.e. each $\dot{\bm{o}}_\text{tip}=\dot{\bm{o}}_\text{base}=0$), and taking $\bm{u}_e=0$, we see that the energy is bounded below and strictly decreases during motion. Using a LaSalle-based argument, we can show that the robot's configuration converges asymptotically to an equilibrium at the minimum of the controlled potential energy with zero velocity.  This outcome is a direct consequence of the interconnected system's passivity. The feedback of passive systems remains passive, and with inherent dissipation, this property guarantees stability \cite{Ortega1998, Ortega2001}. We observe that the passivity of robot and virtual mechanism is a structural property, meaning it is robust to model inaccuracies in the robot mechanics, such as an imprecise inertia matrix. As a consequence, the closed-loop system is robustly stable in the presence of model uncertainties that preserve the passivity of the robot's dynamics.

The stability guarantees from the passivity property of the closed-loop system extend to interactions with a passive environment. This fundamental robustness is what enables the experimental results of Section \ref{sec:Results}, as it ensures closed-loop stability despite real-world uncertainties in both the robot's dynamics and its interactions with the environment.
It is important to note that this stability is not entirely independent of the controller's parameters. The feedback interconnection of the robot and virtual mechanism remains stable only as long as the virtual mechanism is rendered accurately through the joint torques. Therefore, delays and actuator limits affect the maximum achievable stiffness and damping \cite{Pratt1995b, Lawrence1988}.

One interesting feature of the designed system is that the power inflow due to the outer-loop adjustment of the offset can be bounded to $\le k_i (\sigma_\text{tip} |\bm{e}^r_\text{tip}| + \sigma_\text{base} |\bm{e}^r_\text{base}|)$, which grows linearly as the error does rather than quadratically if a linear spring was used.
This can be shown by taking the relevant terms from \eqref{eq:Edot}, substituting the non-zero case of \eqref{eq:Odot} (the zero case being trivial) and using bounds on the magnitude of $\bm{f}_\text{spring}$:
\begin{equation}
\dot{\bm{o}}_\text{tip}^T \bm{f}_\text{spring}(...) 
		+ \dot{\bm{o}}_\text{base}^T \bm{f}_\text{spring}(...) \le k_i(|\bm{e}^r_\text{tip}| \sigma_\text{tip} +|\bm{e}^r_\text{base}| \sigma_\text{base})
        \label{eq:bounds}
\end{equation}
\eqref{eq:bounds} suggests that the injection of energy from the offset adjustment is controlled by the integral gain $k_i$. By selecting a small enough integral gain, 
this energy input can be counteracted by dissipation, leading to local asymptotic stability at equilibrium. Although a formal proof of stability can be developed using Lyapunov analysis \cite{Ortega2023}, we omit it here due to the practical challenges of implementing the outer control loop with computer vision, which introduces further inherent complexity, such as delays and accuracy issues. Instead, we present experimental evidence to demonstrate the stability of the controlled robot.

\section{Experimental Results and Discussion}
\label{sec:Results}

To evaluate the system, we operated on 16 3D printed bone models. 
The surgeon, bone and robot are positioned as shown in Figure \ref{fig:ControllerImplementation}, so that the drill and bone trackers are both visible from the side by the vision sensor.
The bone is raised off the table to imitate the position of the dog leg during surgery, and prevent the drill from impacting the table.

\subsection{Calibration Results}

Calibration of the probe, drill-bit-tip and drill-bit-axis was performed once before starting any of the trials.
The RMS error,
{calculated with \eqref{eq:rms_error},}
for the probe calibration of $\pprobe$ was $\SI{0.225}{mm}$, and for $\ptipdrill$ was $\SI{0.281}{mm}$.
{Using \eqref{eq:axis_rms_err} to compute the RMS error}
for $\adrill$ gave $\SI{0.355}{mm}$.
These are all sub millimetre and represent close to the best accuracy achievable with the Vicra vision sensor.
The drill calibrations using the robot measurements 
{(and the same equations)}
are markedly worse: for $\ptipee$, the RMS error is $\SI{2.81}{mm}$ and for $\aee$, $\SI{2.89}{mm}$.
We expect these error to be worse, due to the kinematic accuracy of the robot, and should not compromise the overall accuracy, as the outer control loop is able to correct for any constant error.

\begin{figure}[htbp]
    \centering
    \includegraphics[width=\columnwidth]{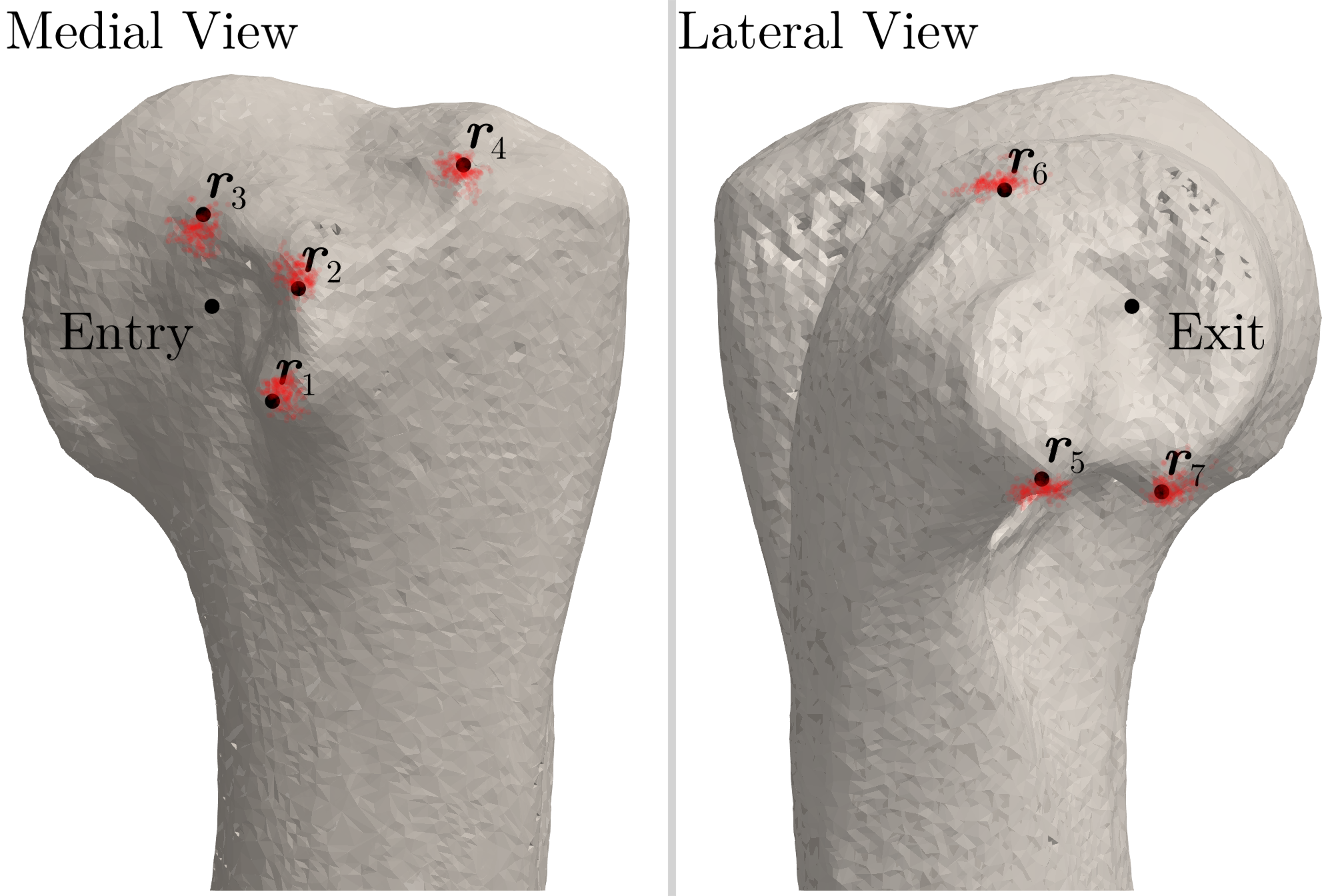}
    \includegraphics[width=\columnwidth]{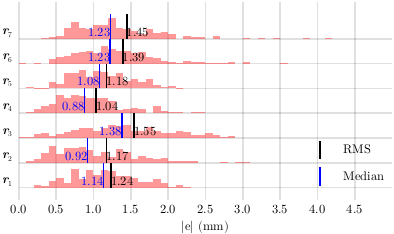}
    \caption{Top: Mesh of the bone formed by segmentation of the CT-Scan with registration points superposed. Each measurement of a registration point is shown as a translucent red dot, having been transformed into the reference frame of the CT-scan by the best-fit transform $T^{sb}$ for that trial. Bottom: histograms showing bone registration error. {The bars represent the distribution of the error between the landmark location and each measurement's location binned at resolution $\SI{0.1}{mm}$}. Data is taken from all trials.}
    \label{fig:RegistrationPoints}
\end{figure}

\begin{figure}[htbp]
    \centering
    \includegraphics[width=\columnwidth]{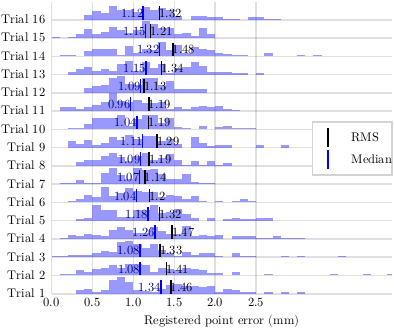}
    \caption{Histograms showing bone registration error for each trial. {The bars represent the distribution of the error between a measurement location and its corresponding landmark binned at resolution $\SI{0.1}{mm}$.}}
    \label{fig:RegistrationTrials}
\end{figure}

For each trial, a new bone registration and a new hand-eye registration were taken.
We present data related to the bone-registration accuracy in Figures \ref{fig:RegistrationPoints} and \ref{fig:RegistrationTrials}.

The histogram in Figure \ref{fig:RegistrationPoints} shows that some points were worse on average than others, such as $\bm{r}_3$.
{Taking the RMS error of only measurements of $\bm{r}_3$ results in an error of}
$\SI{1.55}{mm}$, compared to $\bm{r}_4$ which has an RMS error of only $\SI{1.04}{mm}$. 
Similar data for a range of CT scans could be used to inform decisions about how to choose registration points in the future.  

As seen in Figure \ref{fig:RegistrationTrials}
{computing RMS error on a per trial basis reveals}
error ranging from $\SI{1.13}{mm}$ to $\SI{1.48}{mm}$ over all 16 trials.
This is a potentially significant source of error in drilling, but likely an overestimate of the true bone-registration translation error.
The main expected sources of the registration point RMS error are human error in probe placement and measurement noise. 
If we assume there is no bias in these errors, the resulting registration error over $105$ measurements will be on the order of $1.48/\sqrt{105} = \SI{0.14}{mm}$.
However the zero-bias assumption is not supported by our data, as we can clearly see bias Figure \ref{fig:RegistrationPoints}, as the red clouds of registered points are not centred on the registration points.
We discuss possible improvements to registration in Section \ref{sec:FutureDirections}.

\begin{figure*}[ht]
    \centering
    \includegraphics[width=\columnwidth]{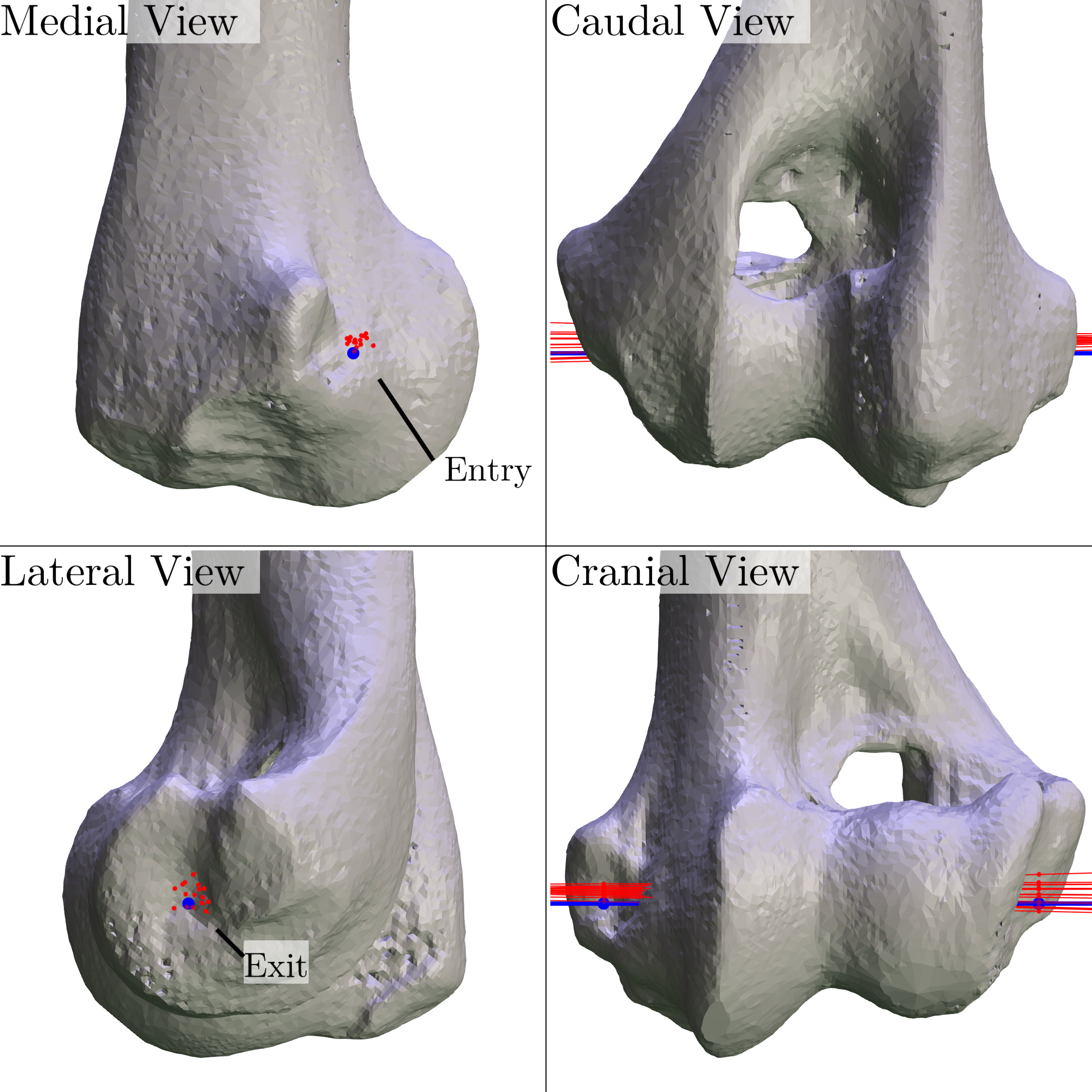}
    \includegraphics[width=\columnwidth]{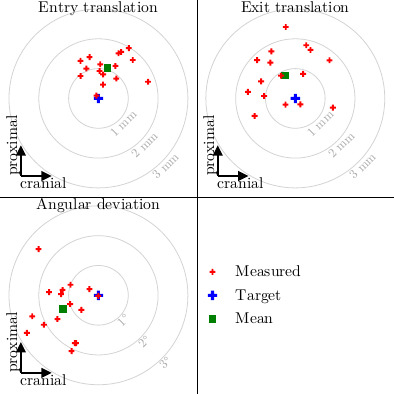}
    \caption{Drilled hole locations, as measured by the vision system. Left: holes shown on the CT scan, with the bone oriented as if the dog were standing. Right: entry/exit/angular errors, all shown from a medial view.}
    \label{fig:Results1}
\end{figure*}

\subsection{Drilling Results}
\label{sec:DrillingResults}

To measure the results, we use the tracked drill bit, shown in Figure \ref{fig:VADAR_setup}. The positions of each end of the drill bit relative to its tracker are determined using the same procedure as detailed in Section \ref{sec:ProbeCalibration}, with resulting RMS errors of $\SI{0.402}{mm}$ and $\num{0.515}{mm}$.

After each trial, we placed the measurement-drill-bit into the drilled hole, and recorded the position of the bone and the drill bit with the vision system.
This gives us an estimate of the trajectories of the drilled holes, which are visualized in Figure \ref{fig:Results1}.
The trajectories are superimposed onto the CT scan, and also projected onto the entry/exit planes.
The angular deviation is also computed.
As these measurements are taken using the vision system, they are dependent upon the bone registration, and the calibration of the measurement drill bit.

\begin{figure}
    \centering
    \includegraphics[width=1\columnwidth]{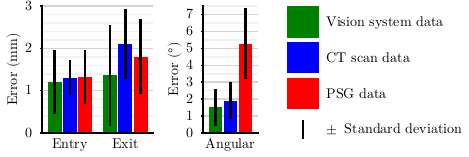}
    \caption{Drilling accuracy results. Compares the mean translation error at entry/exit and the mean angulation error for data gathered by the vision sensor, by the CT scan \cite{Kershaw2025a} and in study \cite{Easter2020} with PSGs. Numerical data is available in Table \ref{tab:results} of the Appendix.}
    \label{fig:results}
\end{figure}

Figure \ref{fig:results} summarizes the results and supports the claim that our system can match the performance of PSGs for transcondylar screw placement, with similar entry/exit accuracy and a significantly improved angular accuracy when compared with PSGs.
This is evidence that a robot-assisted method could provide a suitable alternative for PSGs.
The large angulation error on the PSGs seems incongruous with the comparable exit translations error, however there are several possible explanations.
Firstly, we performed all trials on copies of the same bone, whereas in \cite{Easter2020} multiple different sizes of bone are used: smaller bones can exhibit a larger angulation error for the same translation error.
Secondly, we observe that with the robotic system, translation error at the entry and exit are often in the same direction, whereas for the PSG, the error is more often in opposing directions, allowing for a large angular deviation with a similar entry/exit translation.
For a statistical comparison and further comments on the clinical significance of this work, see \cite{Kershaw2025a}.

The paper \cite{Smith2021} provides a handy summary of entry and destination point accuracy for similar robotic  procedures, to which we compare favourably.
The automated screw-placement robot of \cite{Smith2021} has improved entry accuracy, to which they credit the steel (rather than plastic) tools used---notably however they do not track motion of the bone.
At this level of accuracy, the details of the measurement protocol may significantly affect the results.

To verify the vision system measurements, which cannot capture bone-registration error, we also performed CT scans of the drilled bone models, aligning the new scans with the original scan, and manually annotating the trajectory of the drilled hole.
The process for gathering this data, and a comparative study with PSGs using the same methodology are presented in \cite{Kershaw2025a}.
The results are broadly similar. 
The most significant change is a slight degradation in the measured translation error at the exit.
However, the indicated accuracy of the system is close enough to support our conclusions.

\section{Conclusions and Future Directions}
\label{sec:FutureDirections}

The results of Section \ref{sec:Results} demonstrate that our system 
could provide a compelling alternative to PSGs.
With a comparable level of accuracy, our system present several advantages. 
Soft tissue around the bone must be removed to use a PSG, and residual soft matter causes PSG misalignment, whereas the robotic system requires only a small incision to be made for the drill bit.
A PSG is only suitable for one drill bit size, so either the PSG is only used for the pilot hole, or multiple PSGs must be made and aligned to perform overdrilling to increase the hole size for the fixture \cite{Easter2020}. In contrast, our system allows the same virtual drill guide to be used for overdrilling with any drill bit diameter at no extra cost.
Furthermore, if a procedure requires drilling multiple holes, the robotic system can do so with no additional calibration, registration or hardware required, whereas a unique PSG is required for each intervention.

Our study demonstrates that our robotic system is able to provide the accuracy needed to perform the operation in a less invasive and more flexible way when compared to PSGs.
Further improvements could be made to the system.
These include a better audio-visual interface for the surgeon, to alert to several conditions such as occlusion of the bone-tracker or drill tracker, measured error exceeding certain thresholds and integrator saturation.
Relevant data on operational forces and duration could be extracted to aid in training.
Proper calibration of the robot's kinematics could improve accuracy, therefore reducing (or removing?) burden from the  realtime visual feedback \cite{Kwoh1988}.
The visual feedback mechanism could be more tightly integrated with the virtual mechanism using energy-tank \cite{Ferraguti2015} or power-flow based limits \cite{Shahriari2018}, to provide stronger guarantees about passivity, stability and safety.

There are several challenges to tackle to further improve accuracy.
Forces tangential to the axis of drilling can cause error in 2 ways: by causing fast extension of the springs, too fast for the compensation of vision controller, which only perfectly rejects constant disturbances; and by causing bending between the drill tip and the drill-tracker, which cannot be compensated by the vision controller.
Future studies would need to develop a faster vision controller, also placing trackers closer to the drill bit while avoiding hindering the operation.
High stiffness between the drill mount, drill and drill bit is crucial, which requires replacing plastic mounts with metal parts. 

More recent vision sensors report enhanced accuracy.
These may improve overall system performance, during both calibration and operation.
Contextually, alternative techniques for bone registration could be explored, such as those used in \cite{Brunner2009, Yang2023, Han2024}, where the Ellis pin is screwed into the bone and tracker attached before the CT scan, and its location identified from the scan itself.
A wider exploration of the controller parameters on performance is also needed.
Increasing $k_i$ will makes the integral controller more aggressive, while increasing $k_p$ will increase the stiffness of drill-alignment controller, but both come at the cost of reduced stability margins. 

Finally, while improving the technical capabilities of the system, proper studies are needed to assess the safety of the system, ease of use, and its performance on different bones shapes, and on real bones (rather than PLA).

\appendices

\section{Numerical Drilling Accuracy Results}
Here, we include the numerical results used to generate Figure \ref{fig:results}:
\begin{table}[!h]
    \centering
    \begin{tabular}{r|c|c|c}
         & \specialcell{Mean entry \\ translation \\ (mm)} & \specialcell{Mean exit \\ translation \\ (mm)} & \specialcell{Mean angular \\ deviation ($^{\circ}$)} \\ 
         \hline
        Vision system data & 1.21 ($\pm$ 0.76) & 1.36 ($\pm$ 1.20) & 1.51 ($\pm$ 1.09) \\
        CT scan data \cite{Kershaw2025a} & 1.31 ($\pm$0.43) & 2.1 ($\pm$0.831) & 1.90 ($\pm$1.11)  \\
        PSG data \cite{Easter2020} & 1.33 ($\pm$ 0.64)& 1.80 ($\pm$ 0.89)  & 5.29 ($\pm$ 2.1) \\
    \end{tabular}
    \caption{}
    \label{tab:results}
\end{table}

\section{Linear Approximation of Virtual Drill Dynamics}
\label{ap:poles}
Using the chosen values for $k_{\text{tip}}$, $m_v$ and $b_v$, and $b_{\text{tip}}$ to approximate the dynamics of the virtual drill as a second order system connected to a fixed point (neglecting the base spring and damper which are much smaller) results in:
\begin{align}
m_v\ddot z + (b_v + b_{\text{tip}}) \dot z) + k_{\text{tip}} z &= 0 \\
1.0 \ddot z + 40.5 \dot z + 4000 z &= 0
\end{align}
This gives a pair of poles at $-20.3 \pm 59.9 i$ (to 1 d.p.). 
While oscillatory, these poles decay quickly, bounded by $e^{-20.3t}$.

\bibliographystyle{IEEEtran}
\bibliography{IEEEabrv,library, tempbib}

\vspace{-33pt}
\begin{IEEEbiography}[{\includegraphics[width=1in, height=1.25in, clip, keepaspectratio]{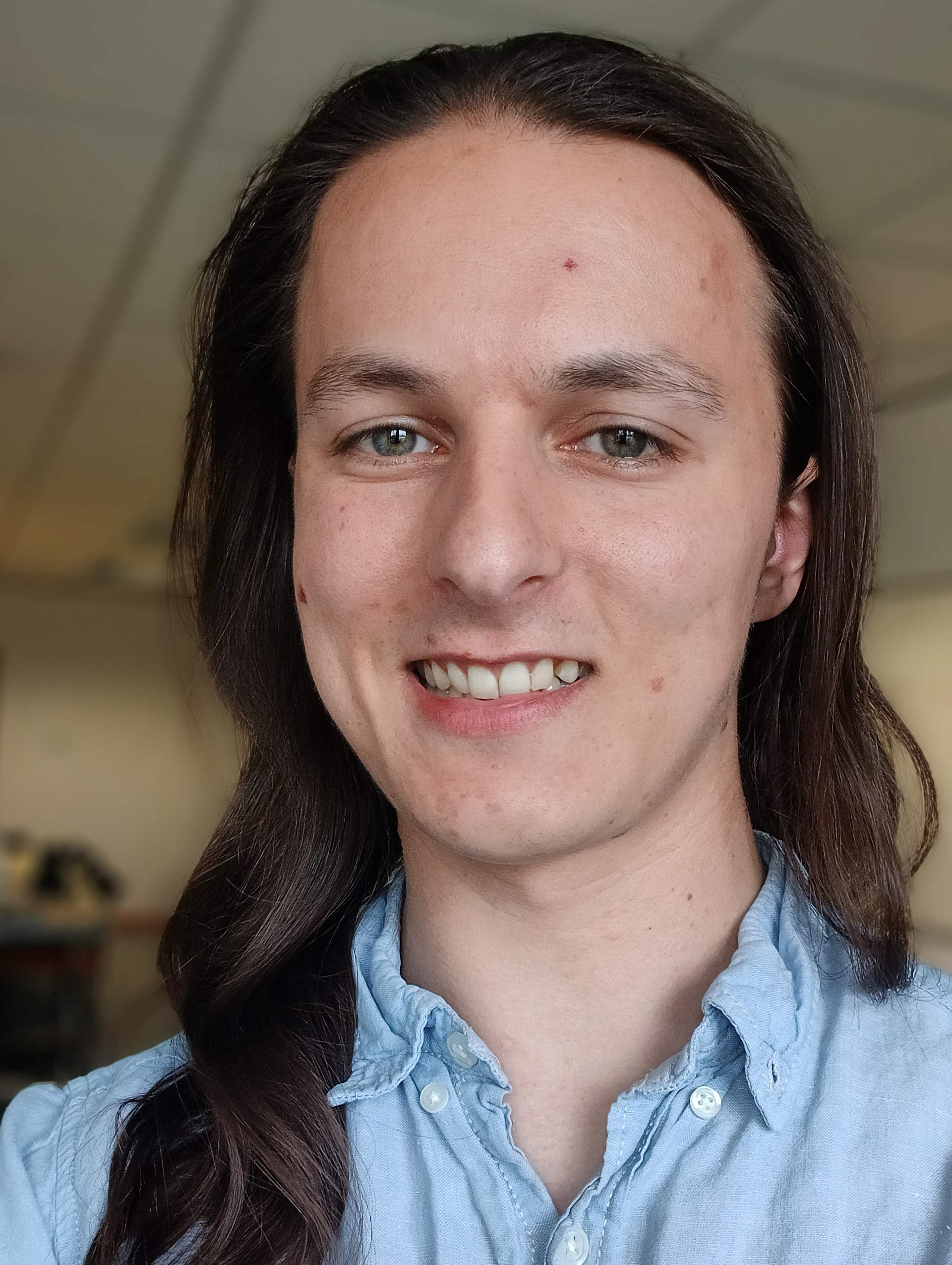}}]{Daniel Larby}
received the MEng degree in control and information engineering in 2020 from the University of Cambridge, UK. He completed his Ph.D. degree in 2025, focussing on robotic control with applications in robotic surgery with the control lab, also in the University of Cambridge, UK.
He now work for Swan EndoSurgical, part of a team working to develop robotic endoluminal surgery solutions.

His research interests include robotics, virtual-mechanism control, passivity based control, impedance control, robotic surgery, and algorithmic differentiation.
\end{IEEEbiography}

\vspace{-33pt}
\begin{IEEEbiography}[{\includegraphics[width=1in, height=1.25in, clip, keepaspectratio]{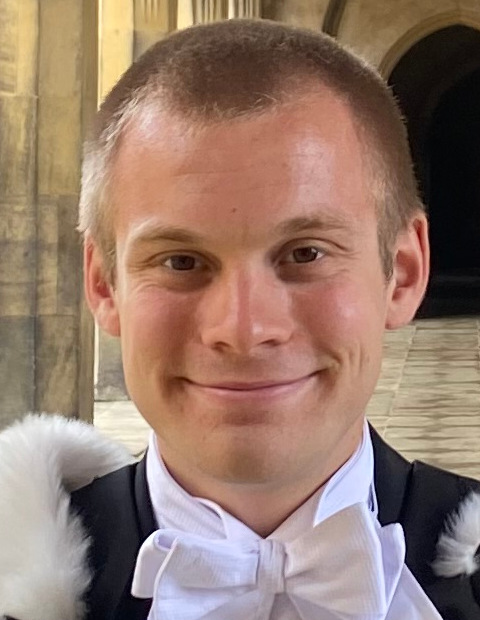}}]{Joshua Kershaw} BA, VetMB, MRCVS, is a Wellcome Clinical Research Fellow (Veterinary Surgery) and 1st year PhD student at the University of Cambridge, supervised by Prof. Matthew Allen in the Surgical Discovery Centre. 

His research interests focus on surgical accuracy and the use of technology in veterinary surgery. 
\end{IEEEbiography}

\vspace{-33pt}
\begin{IEEEbiography}[{\includegraphics[width=1in, height=1.25in, clip, keepaspectratio]{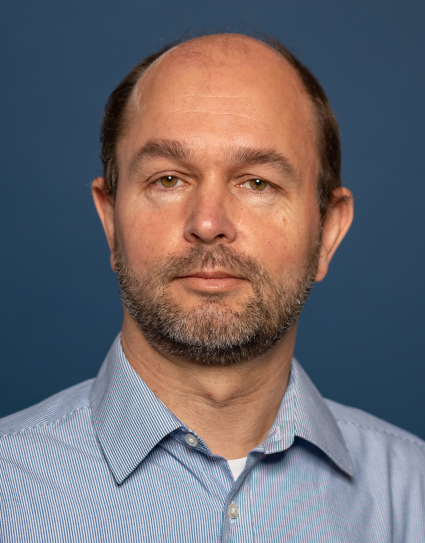}}]{Matthew Allen} trained as a veterinarian, then pursued a PhD in comparative Pathology followed by a postdoctoral fellowship in on the development and optimization of preclinical animal models for orthopaedic conditions. After post-doctoral training at Purdue University, he moved to SUNY Upstate Medical University in Syracuse, where he held joint appointments in the Departments of Orthopedic Surgery and Laboratory Animal Resources. During the 12 years at Syracuse, he mentored a number of graduate students and orthopaedic residents pursuing thesis research. He was recruited to The Ohio State University in March 2008 and directed the Surgical Research Laboratory at the OSU College of Veterinary Medicine from 2008 to 2014, when he was elected Professor of Small Animal Surgery at the University of Cambridge. Matthew’s clinical and research interests lie in the areas of total joint replacement, implantology, orthopaedic oncology, clinical research/clinical trials, surgical navigation and robotics.
\end{IEEEbiography}

\vspace{-33pt}
\begin{IEEEbiography}[{\includegraphics[width=1in, height=1.25in, clip, keepaspectratio]{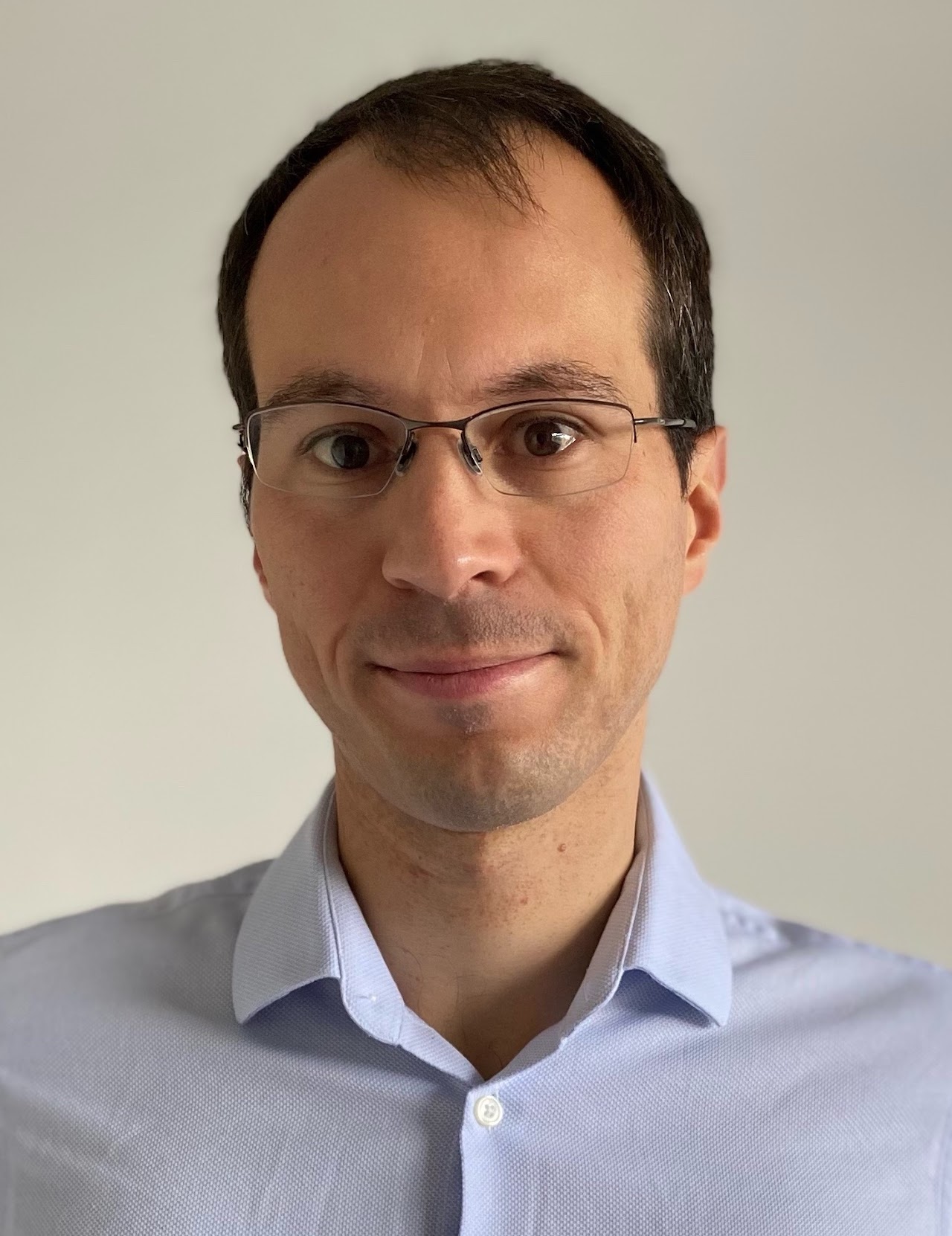}}]{Fulvio Forni}
is a Professor of Control Engineering at the University of Cambridge, where he has been a faculty member since October 2015. He earned his PhD from the University of Rome ‘Tor Vergata’ in 2010. Following his doctorate, he conducted postdoctoral research at the University of Liège in Belgium. Forni’s research interests encompass feedback control and robotics. He received the prestigious IEEE CSS George S. Axelby Outstanding Paper Award in 2020. He is a Director of Studies of Newnham College at Cambridge and serves as a co-investigator for the EPSRC Centre for Doctoral Training in Agrifood Robotics ‘Agriforwards’.
\end{IEEEbiography}

\vfill

\end{document}